\documentclass[sigconf]{acmart}
\AtBeginDocument{%
  }

\copyrightyear{2026}
\acmYear{2026}
\setcopyright{cc}
\setcctype{by}
\acmConference[CIKM '26]{Proceedings of the 35th ACM International Conference on Information and Knowledge Management}{November 07--11, 2026}{Rome, Italy}
\acmBooktitle{Proceedings of the 35th ACM International Conference on Information and Knowledge Management (CIKM '26), November 07--11, 2026, Rome, Italy}
\acmDOI{10.1145/3799682.3840200}
\acmISBN{979-8-4007-2539-5/2026/11}

\usepackage{enumitem}
\usepackage{float}
\usepackage{balance}

\begin{document}

\title{Coverage Under Contradiction: A Benchmark for Negation-Aware Retrieval}

\author{Pasha Abrishamchian}
\affiliation{%
  \institution{Georgetown University}
  \city{Washington}
  \state{DC}
  \country{USA}}
\email{pasha@ir.cs.georgetown.edu}

\author{Eugene Yang}
\affiliation{%
  \institution{Johns Hopkins University}
  \city{Baltimore}
  \state{MD}
  \country{USA}}
\email{eugene.yang@jhu.edu}

\author{Andrew Yates}
\affiliation{%
  \institution{Johns Hopkins University}
  \city{Baltimore}
  \state{MD}
  \country{USA}}
\email{andrew.yates@jhu.edu}

\author{Nazli Goharian}
\affiliation{%
  \institution{Georgetown University}
  \city{Washington}
  \state{DC}
  \country{USA}}
\email{nazli@ir.cs.georgetown.edu}

\author{Ophir Frieder}
\affiliation{%
  \institution{Georgetown University}
  \city{Washington}
  \state{DC}
  \country{USA}}
\email{ophir@ir.cs.georgetown.edu}

\renewcommand{\shortauthors}{Pasha Abrishamchian, Eugene Yang, Andrew Yates, Nazli Goharian, and Ophir Frieder}

\begin{abstract}
Coverage-evaluation benchmarks quantify whether retrieved evidence covers the different pieces of information needed to satisfy an information need. Existing benchmarks mostly consider factual claims that may already be known to pretrained retrieval components, leaving open whether retrievers find corpus evidence when it contradicts those priors. We study how retrievers behave when their parametric knowledge and the information in a corpus disagrees. To do so, we develop a process to augment a nugget-based coverage benchmark with controlled contradictions. For each nugget, we create a negated counterpart and edit its supporting documents to support that negation while preserving the context. Applying this process to CoverageBench, a benchmark measuring information coverage, we create CoverageBench-Negate, spanning six datasets and roughly 63{,}000 original--edited evidence pairs (one per nugget--document link). We attach rich construction metadata to every pair and additionally provide out-of-family LLM judgments over all pairs and a human audit of a sample. In baseline experiments, BM25 separates original from edited near-twin evidence substantially better than dense retrievers (Qwen3 and Jina), which stay near chance. We release the benchmark on Hugging Face Datasets  (nuggets, documents, pairs) together with the construction materials, including the prompts used to create it.
\end{abstract}

\ccsdesc[500]{Information systems~Evaluation of retrieval results}
\ccsdesc[500]{Information systems~Retrieval models and ranking}
\ccsdesc[300]{Computing methodologies~Information extraction}

\keywords{Information retrieval, retrieval-augmented generation, information coverage, benchmark, negation}

\maketitle

\section{Introduction}
\label{sec:introduction}

\begin{figure*}[t!]
  \centering
  \includegraphics[width=\textwidth]{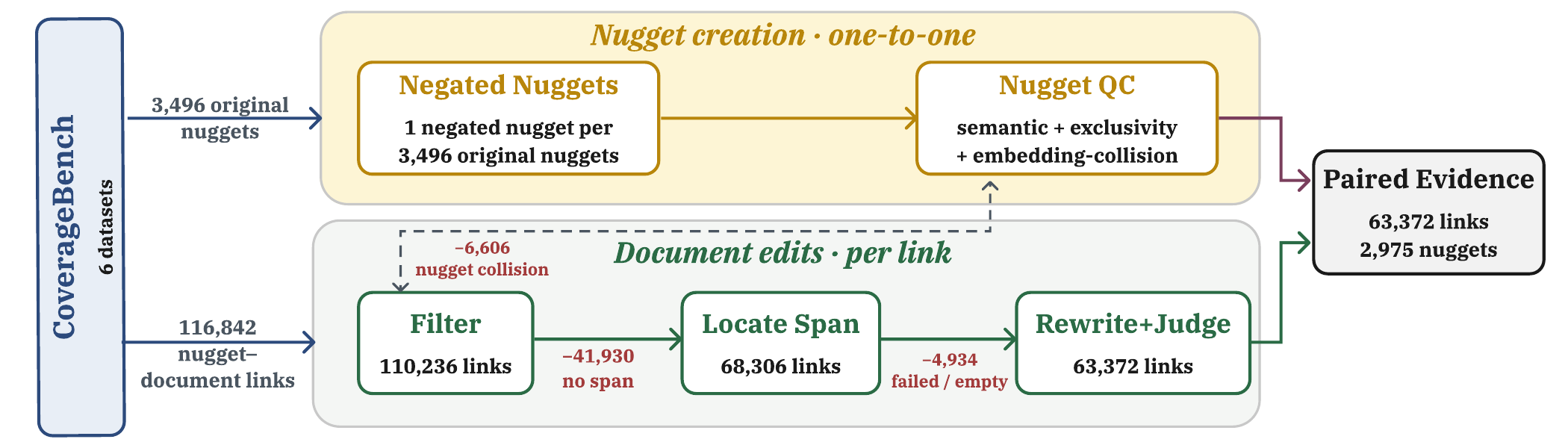}
  \caption{Construction pipeline for CoverageBench-Negate. 
  The nugget lane (top, yellow) produces one negated nugget per original nugget; the document lane (bottom, green) edits one supporting document per nugget--document link, funneling 116{,}842 links down to 63{,}372 released pairs, spanning 2{,}975 nuggets.}
  \label{fig:benchmark-pipeline}
\end{figure*}

Retrieval-augmented generation (RAG) systems rely heavily on the information they retrieve. If the retrieved documents repeat the same point or leave out an important claim, the final answer can be incomplete even when every document is relevant. Retrieval quality, thus, depends not only on finding relevant documents, but on finding a complete set of distinct evidence to support an answer.

Rather than focusing on whether each retrieved document is relevant, coverage evaluation asks whether the retrieved set contains the information needed to answer a query. This information is often represented as \emph{nuggets}: small, atomic claims that a complete answer should cover \cite{pradeepInitialNuggetEvaluation2024,lajewskaGINGERGroundedInformation2025}. Existing coverage benchmarks typically evaluate ordinary historical or world facts that commonly appear in the data used to train language models. 
Meanwhile, modern retrieval pipelines increasingly build on those same language models. 
Many dense retrievers, neural rerankers, and LLM-based query expansion methods inherit, directly or indirectly, the world knowledge absorbed during pretraining. Consequently, they may favor the more familiar version of a claim over the evidence that is actually present in the corpus \cite{sahooNegationNotSemantic2026a}. This raises the question: \textit{when the corpus supports a claim that conflicts with what these language models have seen during pretraining, will the retrieval system still surface that evidence?}

To study this, we develop a general process for augmenting a nugget-based coverage benchmark with controlled contradictions. The process keeps the original topics fixed, and for each nugget, it constructs a negated counterpart. Then, it edits the supporting documents so they support the negation while preserving the surrounding context. Since the topics, entities, and contexts stay fixed, this paired design isolates the polarity flip from confounds such as topic or entity shift. Applying it to CoverageBench \cite{samuelCoverageBenchEvaluatingInformation2026b}, we create CoverageBench-Negate. It covers six datasets: 284 topics and roughly 63{,}000 edited document pairs, spanning every topic and 2,975 nuggets. Each evidence pair is released with quality metadata so researchers can apply strict or permissive filters depending on the diagnostic setting. 

Our contributions are: (1) We develop a reusable process for augmenting nugget-based coverage benchmarks with controlled contradictions to test whether retrievers stay faithful to corpus evidence, and use it to create CoverageBench-Negate; (2) We release the full benchmark on Hugging Face Datasets (negated nuggets, edited supporting documents, a paired evidence set, quality metadata, and our human and LLM-based validations), making it auditable and reusable; and (3) We provide a baseline characterization over retriever-visible top-1000 subsets, showing that BM25 distinguishes original from edited near-twin evidence better than dense retrievers (Qwen3 and Jina).

\section{Background}
\label{sec:background}
Information coverage evaluates a retrieved set of documents by how many distinct pieces of information (aspects) it covers. This underlies recent nugget-based evaluations, where answers are scored for completeness and grounding against nuggets: ARGUE for machine-generated reports \cite{mayfieldEvaluationMachineGeneratedReports2024}, the AutoNuggetizer framework for TREC RAG \cite{pradeepInitialNuggetEvaluation2024}, the Ragnar\"ok RAG evaluation framework \cite{pradeepRagnarokReusableRAG2024}, and nugget-grounded generation pipelines such as GINGER \cite{lajewskaGINGERGroundedInformation2025}. CoverageBench \cite{samuelCoverageBenchEvaluatingInformation2026b} provides a reusable resource for evaluating retrieval by standardizing topics, nuggets, nugget-level relevance labels, document collections, and baseline rankings across multiple retrieval settings. Crucially for a retrieval-focused benchmark, retrieval-side coverage metrics were shown to correlate strongly with the information coverage of downstream RAG responses \cite{samuelRelevanceRelationshipRetrieval2026b}, so evaluating coverage at the retrieval stage is a meaningful proxy for end-to-end RAG behavior.

A related line of work probes whether RAG systems follow the corpus or fall back on parametric knowledge. Closest to our setting, CRUMQs augments RAG evaluation with unanswerable multi-hop queries, testing whether systems invoke parametric knowledge when the required evidence is absent \cite{liuInvestigatingRetrievalAugmentedGeneration2025}. Building on the CoverageBench structure, CoverageBench-Negate poses a complementary question of whether systems still cover corpus-supported claims when those claims conflict with parametric knowledge. CRUMQs probes reliance on outside knowledge by removing support, while we probe it by adding contradicting, corpus-supported evidence.

Work on factual evaluation states that atomic claims must be explicitly supported, not merely topically related \cite{minFActScoreFinegrainedAtomic2023,thorneFEVERLargescaleDataset2018}. Yet dense retrievers, rerankers, and LLM-based query expansion tend to reward semantic similarity, fluent reformulation, and learned associations, and can be brittle under subtle, adversarially constructed perturbations \cite{liuBlackboxAdversarialAttacks2023}. 

Negation is a particularly hard case. Recent analysis shows that dense retrievers may treat two passages as similar because they share a topic even when one states ``X'' and the other ``not X'' \cite{sahooNegationNotSemantic2026a}. Negation-sensitive ranking benchmarks such as NevIR find that retrievers struggle to separate a claim from its negation \cite{wellerNevIRNegationNeural,vandenelsenReproducingNevIRNegation2025,petcuComprehensiveTaxonomyNegation2025}. In a corpus-grounded setting, however, the desired behavior is not to retrieve what is globally likely but what the corpus actually supports \cite{kebirRACRetrievalAugmentedClarification2026}. Testing this cleanly requires paired evidence in which the context stays fixed while only the supported claim changes.

Negation offers a practical way to construct such pairs, flipping the supported claim while leaving the topic, entities, and much of the document intact. Prior work shows that models struggle when negation changes entailment in controlled NLI settings \cite{geigerNeuralNaturalLanguage2020,truongNotAnotherNegation2022,sheScoNeBenchmarkingNegation2023}, precisely the meaning-changing edit we exploit. This mirrors contrast sets, where small label-changing edits test whether a model responds to the changed meaning rather than surface form \cite{gardnerEvaluatingModelsLocal2020}. We use negation as a controlled augmentation rather than a target in itself: each original nugget and supporting document is paired with a negated nugget and an edited document, with construction details in Section~\ref{sec:benchmark}.

\section{Benchmark Construction}
\label{sec:benchmark}

CoverageBench-Negate\footnote{\href{https://huggingface.co/datasets/pashaa/CoverageBench-Negate}{https://huggingface.co/datasets/pashaa/CoverageBench-Negate}.} applies a reusable, negation-based augmentation process to six CoverageBench~\cite{samuelCoverageBenchEvaluatingInformation2026b} datasets: CAsT~2020 \cite{daltonCAsT2020Conversational2021}, NeuCLIR~2024 \cite{lawrieOverviewTREC20242025}, RAG~2024 \cite{pradeepInitialNuggetEvaluation2024}, CRUX-DUC04 \cite{juControlledRetrievalaugmentedContext2026}, CRUX-MultiNews \cite{juControlledRetrievalaugmentedContext2026}, and RAGTIME~2025 \cite{lawrieOverviewTREC20252026a}. We omit Fair Ranking~2022~\cite{ekstrandOverviewTREC20222023} because its nuggets are fairness facets rather than factual claims. During augmentation, we keep the original topics, nuggets, and documents unchanged and add a paired negation layer on top.

Figure~\ref{fig:benchmark-pipeline} illustrates the pipeline. We generate and validate one negated nugget per original nugget, and then edit and validate a supporting document for each surviving link.\footnote{\texttt{Llama-3.3-70B-Instruct} performs nugget negation, span localization, document rewriting, and all LLM-based judgments (sampling $T{=}0.7$, top-$p{=}0.9$, up to 8{,}192 generated tokens); \texttt{Qwen3-Embedding-8B} performs all embedding-based checks.} Since the same model both produces and judges these artifacts, we additionally validate the released pairs with an independent, out-of-family judge and a human audit. The resource consists of the following components:

\begin{itemize}[leftmargin=*, nosep]
  \item \textbf{Topics, original nuggets, and links.} We reuse all 284 CoverageBench topics without modification, and retain the 3,496 nuggets and 116,842 nugget--document support links. A link is a tuple $(t,n,d)$: document $d$ supports nugget $n$ under topic $t$, and 3,234 nuggets have at least one supporting document. Keeping the originals lets the benchmark continue to measure ordinary information coverage alongside the negated coverage.
  \item \textbf{Negated nuggets.} For every nugget $n$, we create one negated nugget $\bar{n}$ that contradicts the original claim while preserving the topic scope, entities, and event context (Section~\ref{sec:negated-nuggets}).
  \item \textbf{Edited documents.} For each support link, we attempt to create an edited document $\bar{d}$ that supports $\bar{n}$ instead of $n$, changing only the target claim while keeping the context intact. Editing is per link rather than per unique document since one source document can support several nuggets. After pre-filtering and failed edits, the 116,842 links yield 63,372 edited documents (Section~\ref{sec:edited-documents}).
  \item \textbf{Paired evidence set.} The main artifact is a link-level paired evidence set of 63,372 pairs spanning every topic, and 2,975 original nuggets. Each pair connects an original topic, nugget, and supporting document with a negated nugget, edited document, and construction metadata.
  \item \textbf{Metadata.} At every stage of the pipeline, we record diagnostics (edit-size and embedding-similarity measurements), validation outcomes 
  , and categorical labels (negation type), so that users can filter the benchmark to their needs. Some checks determine what enters the benchmark; the rest are labels users can filter on (Section~\ref{sec:quality-metadata}).
  
\end{itemize}

\subsection{Negated Nugget Construction}
\label{sec:negated-nuggets}

For each nugget, we prompt the LLM to write a negated version such that the original and the negation cannot both be true. Two LLM checks then refine the result. A single validation call scores three criteria: whether the negation reverses the original meaning, reads naturally as a standalone claim, and preserves the relevant entities. A separate mutual-exclusivity call asks whether the original and negated nuggets could both be true, regenerating the negation from the failed attempt up to three times.
Separately, an embedding check labels each negation by whether it stays on-topic yet meaningfully different from the original and whether it collides with another gold nugget in the same topic. These checks label quality rather than filter, and no nugget is removed at this stage. 

\subsection{Edited Document Construction}
\label{sec:edited-documents}

Editing operates per nugget--document link (116{,}842 in total). We first remove links whose negated nugget was flagged as colliding with another gold nugget in the same topic. Next, the LLM localizes the verbatim span that supports the original nugget; if it finds none, the link is excluded. The model then rewrites only that span so the document supports the negated nugget instead. Deterministic checks verify the edit's length, the absence of synthetic-edit language, and the removal of the original span. A separate LLM judge confirms that the edited document no longer supports the original nugget, does support the negated nugget, and preserves the surrounding context. An edit that fails any check receives one failure-aware retry
. For example, to support the negated nugget ``WBC count often remains elevated even after the infection has cleared,'' the editor rewrites the supporting sentence ``\ldots the child's WBC count will return to normal.'' into ``\ldots the child's WBC count often remains elevated,'' flipping only the target claim while leaving the rest of the document intact. The pipeline yields 63{,}372 evidence pairs (54.2\% of the links) as three exits remove ineligible links: nugget collisions (5.7\%), missing supporting spans (35.9\%), and empty or rejected edits after retry (4.2\%).

\subsection{Metadata}
\label{sec:quality-metadata}
At the nugget creation stage, the LLM tags each negation with one of six types: event negation, outcome inversion, relationship negation, quantitative inversion, antonym substitution, and directional reversal. We also record the cosine similarity between the original and negated nuggets. At the document creation stage, alongside the judge's verdict, we record the edit size and three cosine similarities of the edited document to the negated nugget, the original nugget, and the source document. From these comparisons, we derive a three-tier confidence signal (pass, review, or fail): a pair passes when the edited document is closer to the negated nugget than to the original and stays similar enough to the source.

\subsection{Validation}
\label{sec:benchmark-validation}
We re-judge all 63{,}372 pairs with \texttt{Gemma-4-31B-it} under the same criteria outlined in Section~\ref{sec:edited-documents} and release the full judgments. Gemma confirms all three on 78\% of pairs (87--97\% per criterion). On the nugget layer, Gemma reproduces the negation validation for 82\% of the negated nuggets, but agrees with Llama on mutual exclusivity about 53\% of the time (Cohen's k = 0.05). Judging whether two claims can both be true is subjective, so we record mutual exclusivity as a label rather than filtering based on it. 

Two annotators independently rated a stratified 120-pair sample (60 pass / 30 review / 30 fail by the confidence tiers of Section~\ref{sec:quality-metadata}, all judge-approved) on five criteria mirroring the construction checks. Both confirm the core properties: the negation is a genuine contradiction (90\% and 78\%) and unrelated context is preserved (86\% and 91\%). They are stricter than the automatic judges on the two edit criteria (e.g., the edited document no longer supports the original, 88\% and 58\%), where they also agree less (64\%). The human bar for a flawless pair is thus higher and more variable than the judges'.

\section{Baseline}
\label{sec:baseline}

We provide baseline results for three first-stage retrievers (BM25, \texttt{Qwen3-Embedding-8B}, and \texttt{jina-embeddings-v5-text-small}) to illustrate how CoverageBench-Negate can be used both as an augmented coverage benchmark and as a paired diagnostic resource. The benchmark-style evaluation follows CoverageBench directly. We add the edited documents to the collection and the negated nuggets with their qrels, then score top-20 rankings with $\alpha$-nDCG~\cite{clarkeNoveltyDiversityInformation2008} and Subtopic Recall (StRecall)~\cite{zhaiIndependentRelevanceMethods}. Both are computed at rank~20: $\alpha$-nDCG (standard $\alpha=0.5$) discounts the $j$-th document covering a nugget by $(1-\alpha)^{j-1}$, while StRecall is the fraction of a topic's nuggets covered in the top~20. We use the paired evidence structure to analyze why the dense and lexical retrievers degrade differently.

\subsection{Augmented Coverage Evaluation}
\label{sec:augmented-coverage-evaluation}


\begin{table}[t]
  \centering
  \setlength{\tabcolsep}{2.2pt}
  \caption{Coverage change after augmentation (rank 20, augmented-original); larger negative values indicate larger degradation.}
  \label{tab:augmented-coverage-delta}
  \resizebox{\columnwidth}{!}{%
  \begin{tabular}{lrrrrrr}
    \toprule
    Dataset & \multicolumn{3}{c}{$\Delta\alpha$-nDCG} & \multicolumn{3}{c}{$\Delta$StRecall} \\
    \cmidrule(lr){2-4}\cmidrule(lr){5-7}
     & BM25 & Qwen3 & Jina & BM25 & Qwen3 & Jina \\
    \midrule
    CAsT 2020         & +0.02 & 0     & -0.01 & -0.08 & -0.08 & -0.12 \\
    CRUX-DUC04        & -0.08 & -0.17 & -0.17 & -0.16 & -0.25 & -0.27 \\
    CRUX-MultiNews    & -0.05 & -0.09 & -0.10 & -0.05 & -0.07 & -0.07 \\
    NeuCLIR 2024      & -0.11 & -0.13 & -0.17 & -0.15 & -0.20 & -0.18 \\
    RAG 2024          & -0.13 & -0.27 & -0.25 & -0.23 & -0.24 & -0.20 \\
    RAGTIME           & -0.15 & -0.18 & -0.18 & -0.20 & -0.23 & -0.24 \\
    \bottomrule
  \end{tabular}
  }
\end{table}

Table~\ref{tab:augmented-coverage-delta} reports the per-dataset change from the original CoverageBench evaluation to the augmented evaluation. The augmented benchmark reduces coverage for all three retrievers, but the degradation is larger for the two dense ones. Averaged over datasets, BM25 drops by 0.083 in $\alpha$-nDCG@20 and 0.145 in StRecall@20, while the two dense retrievers degrade almost identically, dropping 0.140--0.146 in $\alpha$-nDCG@20 and 0.180--0.181 in StRecall@20. Since the dense retrievers also start from higher coverage, we compare drops relative to their original scores. The $\alpha$-nDCG gap persists (mean relative drop 16.8\% for BM25 vs.\ 22.3\% and 24.5\% for Qwen3 and Jina), while the StRecall gap nearly closes (21.4\% vs.\ 22.9\% and 23.6\%). The dense retrievers' larger $\alpha$-nDCG sensitivity is thus robust, whereas their larger absolute StRecall drop partly reflects their higher starting coverage.

CAsT is the exception for $\alpha$-nDCG, which remains approximately flat for all three retrievers because some edited documents enter the top-20 as non-redundant gain. StRecall still drops for all three, so we treat CAsT as metric-specific rather than evidence against the degradation pattern. Overall, these results indicate that, in most cases, the retrieval models do not rank documents supporting negated claims the way they rank the originals.

\subsection{Paired Score-Gap Diagnostic}
\label{sec:score-gap-evaluation}

For each retriever, query, and original--edited document pair, we compare the raw score assigned to the original document, $s_{orig}$, with the raw score assigned to the edited negation-supporting document, $s_{edit}$. We define the score gap as $\Delta_s = s_{edit} - s_{orig}$. A negative value means the retriever prefers the original document, while a positive value means it prefers the negated document. This analysis uses retriever-specific subsets of the paired evidence set (28,952 pairs for BM25, 47,054 for Qwen3, and 43,942 for Jina) where both $s_{orig}$ and $s_{edit}$ are available in the top-1000 retrieved documents. The two scores are compared within the same retriever and query setting, so the sign of $\Delta_s$ gives a direct pairwise preference signal, though the raw magnitudes are comparable only within a retriever.

We stratify pairs by the document edit size (character-level difference between the original and edited documents) and assign each pair to a bucket ranging from tiny (<2\% change) to vast  (>25\% change) edits. Small edits are the cleanest test since most wording and topic remain unchanged. Any shift in a retriever's scores can then only come from the flipped claim.


\begin{table}[t]
  \centering
  \small
  \setlength{\tabcolsep}{2.7pt}
  \caption{Score-gap by edit length over retriever-specific top-1000 subsets. \textsc{Pref.} is the percentage of pairs where the retriever scores the original document above its negated counterpart; Med. is median $|\Delta_s|$.}
  \label{tab:score-gap-edit-length}
  \resizebox{\columnwidth}{!}{%
  \begin{tabular}{llrrrrrr}
    \toprule
    Retriever & Metric & Tiny & Small & Medium & Large & Vast & All \\
    \midrule
    BM25  & \textsc{Pref.} & 83.90 & 72.10 & 65.70 & 61.30 & 61.40 & 68.80 \\
    BM25  & Med.           & 0.13 & 0.16 & 0.24 & 0.44 & 0.63 & 0.24 \\
    \midrule
    Qwen3 & \textsc{Pref.} & 58.10 & 54.00 & 54.20 & 55.20 & 60.50 & 56.00 \\
    Qwen3 & Med.           & 0.01 & 0.01 & 0.01 & 0.02 & 0.02 & 0.01 \\
    \midrule
    Jina  & \textsc{Pref.} & 56.30 & 50.90 & 49.90 & 50.90 & 55.20 & 52.20 \\
    Jina  & Med.           & 0.00 & 0.01 & 0.01 & 0.01 & 0.02 & 0.01 \\
    \bottomrule
  \end{tabular}
  }
\end{table}

Table~\ref{tab:score-gap-edit-length} explains the coverage deltas. BM25 prefers the original in about 84\% of tiny-edit pairs, decreasing to about 61\% as edits grow. 
In contrast, the two dense retrievers stay near chance across every bucket (Qwen3 54--61\%, Jina 49--56\%), with near-zero median gaps on their scale, so they cannot reliably tell the original from its negation. The contrast is sharpest on near-twin pairs, where BM25 discriminates far better than either dense retriever. The same pattern carries over to the actual rankings a downstream system consumes. 
When both documents fall in the same with-negated top-1000 list (26,824 BM25, 44,034 Qwen3, 42,140 Jina pairs), BM25 ranks the original above the edit in 83\% of tiny-edit pairs, falling to 57\% for vast edits, while Qwen3 (roughly 51--55\%) and Jina (48--55\%) stay near chance across the sub-25\% buckets.\footnote{Full per-bucket and per-dataset score-gap and rank-level numbers are released with the benchmark on Hugging Face.}
\section{Conclusion}
\label{sec:conclusion}

CoverageBench-Negate tests whether first-stage retrievers cover corpus evidence that contradicts pretrained expectations. Its negated nuggets, edited documents, paired evidence, and quality metadata enable studies of corpus faithfulness under controlled contradiction. Our initial characterization separates lexical and dense retrievers, making this a useful diagnostic for coverage-oriented retrieval.

Its premise is that a negated claim conflicts with parametric knowledge, but we do not verify this conflict directly, and we do not analyze whether retained links differ systematically from those excluded. Independent edits may introduce higher-order contradictions. Construction and quality checks are LLM-based, and we mitigate this with an independent out-of-family judge and a human audit.

\section*{GenAI Usage Disclosure}
Generative AI is central to the construction of CoverageBench-Negate. \texttt{Llama-3.3-70B-Instruct} generates every negated nugget, localizes the supporting span, rewrites each edited document, and performs all LLM-based quality checks (semantic validation, mutual exclusivity, and the document judge). \texttt{Qwen3-Embedding-8B} produces all embedding-based checks and confidence tiers, and \texttt{Gemma-4-31B-it} serves as the independent, out-of-family validation judge. The released negated nuggets, edited documents, and paired evidence set are therefore LLM-generated and LLM-judged, as detailed in Section~\ref{sec:benchmark}. The authors additionally used generative AI tools to assist with writing and editing, code debugging, and figure creation. The authors reviewed and take full responsibility for all content and analyses presented in this paper.

\bibliographystyle{ACM-Reference-Format}
\balance
\bibliography{sample-base,custom}

\end{document}